\documentclass[12pt]{article}
\usepackage{amsfonts}
\usepackage{latexsym}
\title{Asymptotics of eigenvalues of the Aharonov-Bohm operator with a 
strong  $\delta$-interaction on a loop}
\newlength{\dinwidth}
\newlength{\dinmargin}
\def\ben{\begin{displaymath}}
\def\een{\end{displaymath}}
\def\beq{\begin{equation}}
\def\eeq{\end{equation}}
\def\beqs{\begin{displaymath}}
\def\eeqs{\end{displaymath}}
\def\beqn{\begin{eqnarray}}
\def\eeqn{\end{eqnarray}}

\def\R{\mathbb {R}}
\def\N{\mathbb{N}}

\newtheorem{thm}{Theorem}[section]

\newtheorem{prop}{Proposition}[section]
\newtheorem{coro}{Corollary}[section]
\newtheorem{lem}{Lemma}[section]

 \def \R {\Bbb R}
 \def \N {\Bbb N }

\begin{document}
\title{Asymptotics of eigenvalues of the Aharonov-Bohm operator with a 
strong  $\delta$-interaction on a loop}
\author{G.~Honnouvo $^{1,2,3}$ and M. N.~Hounkonnou $^{1,2}$}
\date{}
\maketitle
\begin{quote}{\small \quad ${}^{1}$ International Chair in Mathematical Physics and Applications (ICMPA)\\
\phantom{e)x}072 BP.: 50 Cotonou, Benin\\
\phantom{e)x} ${}^{2}$  Unit\'e de Recherche en Physique Th\'eorique (URPT)\\
\phantom{e)x}Institut de Math\'ematiques
et de Sciences  Physiques (IMSP) \\
\phantom{e)x}01 B.P. 2628 Porto-Novo, Benin \\
\phantom{e)x}${}^{3}$ Department of Mathematics and Statistics, Concordia University, \\ 
\phantom{e)x}7141 Sherbrooke Street West, Montreal, Quebec H4B 1R6, Canada \\
\rm \phantom{e)x}g$_{-}$honnouvo@yahoo.fr, norbert${}_{-}$hounkonnou@cipma.net}

\vspace{8mm}

\noindent {\small \begin{abstract}
We investigate the two-dimensional Aharonov-Bohm operator
 $H_{c_0,\beta} = {(-i\nabla -A)}^{2}-\beta\delta(.-\Gamma),$ where 
$\Gamma$ is a smooth loop and $A$ is the vector potential which corresponds to 
Aharonov-Bohm potential. The asymptotics of negative eigenvalues of 
$H_{c_0,\beta}$ for $\beta \longrightarrow +\infty$ is found. We also prove that 
for large enough positive value of $\beta$ the system exhibits persistent currents.
\end{abstract}}
\end{quote}

\section{Introduction}
In the presence of a static magnetic field, a single isolated 
normal-metal loop is predicted to carry an equilibrium current\cite{BIL}, which 
is periodic in the magnetic flux $\Phi$ threading the loop. This current 
arises due to the boundary conditions \cite{BY}
imposed by the doubly connected nature of the loop. As a consequence of 
these boundary conditions, the free energy $E$ and the thermodynamic 
current $I(\Phi) = \frac{\partial E}{\partial \Phi}$ are periodic in 
$\Phi,$ with a fundamental period $\Phi_0 = \hbar /e .$ 
In recent papers \cite{EY1} '\cite{EY2} Exner and Yoshitomi have 
derived an asymptotic formula showing that if the $\delta-$coupling is strong 
or in a homogeneous magnetic fied $B$ perpendicular to the plane, the 
negative eigenvalues approach those of the ideal model in which the 
geometry of $\Gamma$ is taken into account by means of an effective 
curvature-induced potential.
The pupose of this paper is to ask a similar question in the situation 
when the electron is subject to a Bohm-Aharonov potential. We are going to derive an analogous asymptotic formula where the presence of the magnetic field
 is taken into account via the boundary conditions specifying the
 domain of the comparison operator as in \cite{EY2}.
As a consequence of this result, we prove that the system exhibits persistent 
currents.

\section{The model and the results}
In this section, we study  the Aharonov-Bohm operator in 
$L^2(\R^2)$ with an attractive $\delta$-interaction applied to a loop. We use 
the gauge fied $A = c_0\bigg(\frac{-y}{{x^2+y^2}}; 
\frac{x}{{x^2+y^2}}    \bigg).$
Let $ \Gamma :[0,L] \ni s \mapsto (\Gamma_1(s), \Gamma_2(s)) \in 
\R^2$ be the closed counter-clockwise $C^4$ Jordan curve which is parametrized 
by its arc length. Given $\beta >0$ and $c_0 \in ]0,1[,$ we define the 
quadratic form

\begin{eqnarray}
q_{c_0,\beta}(f;f)=||(-i\partial _x + \frac{c_0y}{{x^2+y^2}} 
)f||^2_{L^2(\R^2)} + ||(-i\partial _y - \frac{c_0x}{{x^2+y^2}})f||^2_{L^2(\R^2)}    
-\beta\int_\Gamma |f(x)|^2ds \nonumber
\end{eqnarray}

with the domain $H^1(\R^2)$, where $\partial_x \equiv 
\frac{\partial}{\partial_x}$, and the norm refers to $L^2(\R^2).$

Let us denote by $H_{c_0, \beta}$ the self-adjoint operator associated 
to the form $q_{c_0, \beta}(\:,\:)$:

$$ H_{c_0, \beta} = {(-i\nabla -A)}^{2}-\beta\delta(.-\Gamma).$$
Our main goal is to study, as in \cite{EY2}, the asymptotic behaviour of the negative 
eigenvalues of $H_{c_0, \beta}$
as $\beta \longrightarrow +\infty.$ 

Let $\gamma :\R   \longrightarrow \R$
be the signed curvature of $\Gamma$ , i.e.
$$ \gamma(s):= \bigg( 
\Gamma_1''\Gamma_2'-\Gamma_2''\Gamma_1'\bigg)(s).$$

Next we need a comparison operator on the cuve
\begin{eqnarray}
S_{c_0} = -\frac{d^2}{ds^2}- \frac{1}{4}\gamma(s)^2\:\: \mbox{in}\:\: L^2\bigg((0; L)\bigg),
\end{eqnarray}

with the domain 
\begin{eqnarray}
P_{c_0} = \{ u\in H^2(]0; L[); u^{(k)}(L) = u^{(k)}(0);\:k= 1,2 \}.
\end{eqnarray}
For $j\in\N,$ we denote by $\mu_j(c_0)$ the $j^{th}$ eigenvalue of the 
operator $S_{c_0}$ counted with multiplicity. This allows us to formulate 
our main result, and his proof follows in the same way as in \cite{EY2}:

\smallskip\begin{thm}\label{t1} 
Let $n$ be an arbitrary integer and $I$ be a nonempty compact subset of $]0,1[$. Then  there exists 
$\beta (n, I)$ such that $\# \bigg\{ \sigma_{d}(H_{c_0,\beta })\cap 
]-\infty, 0[ \bigg\}\geq n$  for $\beta\geq \beta(n,I)$ and $c_0 \in I.$ 

For $\beta\geq \beta(n, I)$ and $c_0 \in I$ we denote by 
$\lambda_n({c_0,\beta})$ the $n^{th}$ eigenvalue of $H_{c_0, \beta}$ counted with 
multiplicity.

Then $\lambda_n({c_0, \beta })$ admits an asymptotic expansion of the 
form

$\lambda_n({c_0, \beta }) = -\frac{1}{4}\beta^2 + \mu_n(c_0) + \mathcal{O} 
(\beta^{-1}\ln \beta)$ as $\beta\rightarrow +\infty$; where the error term 
is uniform with respect to $c_0\in I.$
\end{thm}\smallskip

The existence of persistent currents is given by the  
consequence of the following result.

\smallskip\begin{coro}\label{c1} 
Let $n\in \N.$ Then there exists a constant $\beta_1(n, 
I)>0$ such that the function $\lambda_n({.,\beta})$ is not constant for 
$\beta >\beta_1(n, I) .$
\end{coro}\smallskip

Since the spectral properties of $H_{c_0,\beta}$ are cleary invariant 
with respect to Euclidean transformation of the plane, we may assume 
without any loss of generality that the curve $\Gamma$ parametrizes in the 
following way:

$\Gamma_1(s) = \Gamma_1(0) + \int_0^s \cos H(t) dt \qquad \Gamma_2(s) = 
\Gamma_2(0) + \int_0^s \sin H(t) dt$

where $H(t) \equiv -\int_0^t \gamma (u)du .$ Le $\Psi_a$ be the map

$ \Psi_a :[0,L)\times(-a,a)\ni(s,u)\mapsto (\Gamma_1(s)-u\Gamma_2'(s), 
\Gamma_2(s)+u\Gamma_1'(s) ) \in \R^2.$

From \cite{EY1} we know that there exists $a_1>0$ such that the map 
$\Psi_a$ is injective for all $a\in (0,a_1].$ We thus fix $a\in(0,a_1)$ 
and denote by $\Sigma_a$ the strip of width $2a$ enclosing $\Gamma.$

$$\Sigma_a \equiv \Psi_a([0,L)\times (-a,a)).$$

Then the set $\R^2/{\Sigma_a}$ consists of two connected components 
which we denote by $\wedge_a ^{in}$ and $\wedge_a ^{out},$ where the 
interior one, $\wedge_a ^{in},$ is compact. We define a pair of quadratic 
forms,

\begin{eqnarray}
q_{c_0 ,a, \beta} ^{\pm}(f;f)=||(-i\partial _x + \frac{c_0y}{{x^2+y^2}} 
)f||^2_{L^2(\Sigma_a)} + ||(-i\partial _y - 
\frac{c_0x}{{x^2+y^2}})f||^2_{L^2(\Sigma_a)}    -\beta\int_\Gamma |f(x)|^2ds \nonumber
\end{eqnarray}

which are given by the same expression but differ by their domains, the 
latter in $H_0^1(\Sigma_a)$ for $q_{c_0 ,a, \beta} ^{+}$ and $H^1(\Sigma_a)$ for 
$q_{c_0 ,a, \beta} ^{-}.$ Furthermore, we introduce the quadratic forms

\begin{eqnarray}
e_{c_0, a} ^{\pm}(f;f)=||(-i\partial _x + \frac{c_0y}{{x^2+y^2}} 
)f||^2_{L^2(\wedge_a ^j)} + ||(-i\partial _y - 
\frac{c_0x}{{x^2+y^2}})f||^2_{L^2(\wedge_a ^j)} 
\end{eqnarray}

for $j= out,\: in,$ with the domain $H_0^1(\wedge_a ^j)$ and  
$H^1(\wedge_a ^j)$ corresponding to the $\pm$ sign respectively. Let $L_{c_0 , 
a, \beta}^{\pm},\: E_{c_0, a}^{out, \pm}$ and $ E_{c_0, a}^{in, \pm}$ 
be the self-adjoint operators associated with the forms $q_{c_0,a, \beta 
}^{\pm}, \: e_{c_0, a} ^{out,\pm}$ and $e_{c_0, a} ^{in,\pm},$ 
respectively.

As in \cite{EY1} we are going to use the Dirichlet-Neumann bracketing 
with additional boundary conditions at the boundary of $\Sigma_a.$ One can easily see this by comparing the form domains of the 
involved operators, cf[\cite{EY2} or (\cite{rs}, thm XIII.2)]. We get

\begin{eqnarray}\label{ok}
E_{c_0, a}^{in,-}\oplus L_{c_0 , a, \beta}^{-} \oplus E_{c_0, 
a}^{out,-} \leq H_{c_0,a}\leq E_{c_0, a}^{in,+}\oplus L_{c_0 , a, \beta}^{+} 
\oplus E_{c_0, a}^{out,+}    
\end{eqnarray}
with the decomposed estimating operators in $L^2(\R^2) = L^2(\wedge_a 
^{in})\oplus L^2(\Sigma_a )\oplus L^2(\wedge_a ^{out}).$ In order to 
assess the negative eigenvalues of $H_{c_0, \beta},$ it suffices to 
consider those of $L_{c_0 , a, \beta}^{+}$ and $L_{c_0 , a, \beta}^{-},$ 
because the other operators involved in $(\ref{ok})$ are positive. Since the loop 
is smooth, we can pass inside $\Sigma_a$ to the natural curvilinear 
coordinates. We state

$$(U_af)(s,u) = {(1 + u\gamma(s))}^{1/2}f(\Psi_a(s,u))\qquad 
\mbox{for}\: f\in L^2(\Sigma_a )$$

which defines the unitary operator $U_a$ from $L^2(\Sigma_a )$ to 
$L^2((0,L )\times(-a,a)).$ To express the estimating operators in the new 
variables, we introduce
\begin{eqnarray*}
{\cal{Q}} _a^+  &=& \bigg\{ \psi\in H^1((0,L )\times(-a,a));\:\psi(L,.) 
=\psi(0,.)\quad\:on\:\quad (-a,a);\nonumber\\
 & & \psi(.,a) =\psi(.,-a)\:\quad on \quad\: (0,L)  \bigg\}
\end{eqnarray*}
\begin{eqnarray*}
{\cal{Q}} _a^- &=&  \bigg\{ \psi\in H^1((0,L )\times(-a,a));\quad \psi(L,.) 
=\psi(0,.)\quad \: on\: \quad(-a,a) \bigg\}  
\end{eqnarray*}

and define the quadratic forms

\begin{eqnarray}\label{ba4}
b_{c_0,a,\beta}^\pm [g] &=&\int_0^L \int_{-a}^a {(1 + 
u\gamma(s))}^{-2}|\partial_s g|^2 duds + \int_0^L \int_{-a}^a|\partial_u g|^2 \nonumber 
\\
&+& \int_0^L \int_{-a}^a V(s,u) |g|^2 dsdu - \beta \int_0^L|g(s,0)|^2ds 
\nonumber \\
& -& \frac{b_\pm}{2}\int_0^L \frac{\gamma (s)}{1 + a \gamma 
(s)}|g(s,a)|^2ds + \frac{b_\pm}{2}\int_0^L \frac{\gamma (s)}{1 - a \gamma 
(s)}|g(s,-a)|^2ds \nonumber \\
&+& c_0^2\int_0^L \int_{-a}^a \theta(s,u)|g|^2duds \nonumber \\
&+& 2c_0 Im \int_0^L \int_{-a}^a \theta (s,u)(\Gamma_2  + u 
\Gamma_1')\bigg({(1+ u\gamma)}^{-1}\cos H \overline{g}\partial_s g - \sin H 
\overline{g}\partial_u g  \bigg) duds \nonumber \\
&-& 2c_0 Im \int_0^L \int_{-a}^a \theta (s,u)(\Gamma_1  - u 
\Gamma_2')\bigg({(1+ u\gamma)}^{-1}\sin H \overline{g}\partial_s g + \cos H 
\overline{g}\partial_u g  \bigg) duds\nonumber\\
\end{eqnarray}

on ${\cal{Q}}_a^\pm $ respectively, where

$V(s,u) = \frac{1}{2}{(1 + u\gamma (s))}^{-3}u \gamma(s)'' -\frac{5}{4}{(1 + u\gamma 
(s))}^{-4}u^2\gamma'(s)^2 -\frac{1}{4}{(1 + u\gamma (s))}^{-2}\gamma (s)^2 , $ 

$\theta (s,u) = {\bigg( \Gamma_1^2(s) + \Gamma_2^2(s) + u^2 - 2u 
\big(\Gamma_1(s)\Gamma_2'(s) -\Gamma_2(s)\Gamma_1'(s) \big) \bigg) }^{-1}$

$b_+=0$ and $b_- = 1.$

Let $D_{c_0, a,\beta}^{\pm}$ be the self-adjoint operators associated 
with the forms 
$b_{c_0,a,\beta}^\pm $, respectively. By analogy with \cite{EY1}, we 
get the following result.

\smallskip\begin{lem}\label{1} 
$U_a D_{c_0, a,\beta}^{\pm}U_a = L_{c_0, a,\beta}^{\pm}.$
\end{lem}\smallskip

In order to eliminate the coefficients of $\overline{g}\partial_s g$ 
and $\overline{g}\partial_u g$ in (\ref{ba4}) modulo small erros, we 
employ the following unitary operator

\begin{eqnarray}\label{up}
(M_{c_0}h)(s,u) = \exp [iK(s,u)]h(s,u).
\end{eqnarray}

Replacing $M_{c_0}h$ in (\ref{ba4}), it becomes:

\begin{eqnarray}\label{ba1}
c_{c_0,a,\beta}^\pm [g] &=&\int_0^L \int_{-a}^a {(1 + 
u\gamma(s))}^{-2} |g_s|^2  duds
+ \int_0^L \int_{-a}^a | g_u|^2duds \nonumber 
\\
&-&  \beta \int_0^L|g(s,0)|^2ds -\frac{b_\pm}{2}\int_0^L \frac{\gamma (s)}{1 + a \gamma 
(s)}|g(s,a)|^2ds + \frac{b_\pm}{2}\int_0^L \frac{\gamma (s)}{1 - a \gamma 
(s)}|g(s,-a)|^2ds \nonumber \\
&+&\int_0^L \int_{-a}^a \bigg( \theta(s,u)c_0^2 + {(1 +u\gamma(s))}^{-2}K_s ^2+K_u ^2 +V(s,u)\nonumber\\ \nonumber
&+&2c_0\Omega_1(s,u)K_s-2c_0\Omega_2(s,u)K_s\bigg)|g|^2duds \nonumber \\
&+& 2Im \int_0^L \int_{-a}^a\bigg( c_0\Omega_1(s,u)+{(1 + u\gamma(s))}^{-2}K_s\bigg)\overline{g}g_s  duds \nonumber \\
&-& 2 Im \int_0^L \int_{-a}^a \bigg(c_0\Omega_2(s,u)-K_u\bigg)\overline{g}g_u  duds 
\end{eqnarray}

where 
\begin{eqnarray}
\Omega_1(s,u) = \theta (s,u)\bigg(\Gamma_2 \cos H  
-\Gamma_1\sin H + u\bigg){(1+ u\gamma)}^{-1},\\
\Omega_2(s,u)=\theta (s,u)\bigg(\Gamma_1 \cos H 
+\Gamma_2\sin H \bigg){(1+ u\gamma)}^{-1}\\
K_s=\frac{\partial_s K(s,u)}{\partial s},\:\:K_u=\frac{\partial_u K(s,u)}{\partial u}\:\:g_s=\frac{\partial_s g(s,u)}{\partial s},\:\:g_u=\frac{\partial_u g(s,u)}{\partial u}
\end{eqnarray}

To eliminate the coefficients of $\overline{g}\partial_u g$ in $c_{c_0,a,\beta}^\pm [g] $, we have the following differential equation:

 \begin{eqnarray}
\frac{\partial K(s,u)}{\partial_u}=c_0\Omega_2(s,u) 
\end{eqnarray}

and then, we have

\begin{eqnarray}
K(s,u)=\int_0^u c_0\Omega_2(s,v) dv.
\end{eqnarray}

This form of $K,$ reduces (\ref{ba1}) to:

\begin{eqnarray}\label{b}
\tilde b_{c_0,a,\beta}^\pm [g] &=&\int_0^L \int_{-a}^a {(1 + 
u\gamma(s))}^{-2} |g_s|^2  duds
+ \int_0^L \int_{-a}^a | g_u|^2duds \nonumber 
\\
&-&  \beta \int_0^L|g(s,0)|^2ds -\frac{b_\pm}{2}\int_0^L \frac{\gamma (s)}{1 + a \gamma 
(s)}|g(s,a)|^2ds + \frac{b_\pm}{2}\int_0^L \frac{\gamma (s)}{1 - a \gamma 
(s)}|g(s,-a)|^2ds \nonumber \\
&+&\int_0^L \int_{-a}^a \bigg( \theta(s,u)c_0^2 + {(1 +u\gamma(s))}^{-2}K_s ^2+K_u ^2 +V(s,u)\nonumber\\ 
&+&2c_0\Omega_1(s,u)K_s-2c_0\Omega_2(s,u)K_s\bigg)|g|^2duds \nonumber \\
&+& 2\int_0^L \int_{-a}^a\bigg( c_0\Omega_1(s,u)+{(1 + u\gamma(s))}^{-2}K_s\bigg)Im\overline{g}g_s  duds, 
\end{eqnarray}

for $g\in  {\cal{Q}} _a^\pm ,$ respectively.

Let us remark that because of the properties of the curve $\Gamma$, we have $\Omega_2(0, u)= \Omega_2(L, u)\:\: \forall u\in (-a,a)$. So the domains ${\cal{Q }}_a^{\pm}$ are not changed under the unitary operator $M_{c_0}.$

Let $\tilde D_{c_0,a,\beta}$ be the self-adjoint operators associated with the 
forms $\tilde b_{c_0,a,\beta}^\pm ,$ respectively. We have the 
following result

\smallskip\begin{lem}\label{2} 
$M_{c_0}^* D_{c_0, a,\beta}^{\pm}M_{c_0} = \tilde D_{c_0, 
a,\beta}^{\pm}.$
\end{lem}\smallskip

 In the estimation of the $\tilde D_{c_0, a,\beta}^{\pm}$, let us use 
the same notation as  in \cite{EY2}:

$\gamma_+ =   \displaystyle{\max_{[0,L]}}| \gamma(.)|$

$N_{c_0}(a)=\displaystyle{ \max_{(s,u)\in [0,L]\times[-a,a]}} 2|c_0\Omega_1(s,u)+{(1 + u\gamma(s))}^{-2}K_s|$
 
and

$M_{c_0}(a) :=\displaystyle{ \max_{(s,u)\in [0,L]\times[-a,a]}} 
|W_{c_0}(s,u)+\frac{1}{4}\gamma (s)^2 |;$ where

\begin{eqnarray}\label{ab}
W_{c_0}(s,u)&=&\theta(s,u)c_0^2 + {(1 +u\gamma(s))}^{-2}K_s ^2+K_u^2+V(s,u)\nonumber\\
&+&2c_0\bigg(\Omega_1(s,u)K_s-\Omega_2(s,u)K_s\bigg).
\end{eqnarray}

Since $c_0 \in I$ and $I$ is a compact interval, then there exists 
$T$ such that 

$N_{c_0}(a) + M_{c_0}(a) \leq Ta \qquad $ for $0< 
a<\frac{1}{2\gamma_+}$ and  $ c_0 \in I,$
where $T$ is independent of $a$ and $c_0$. For fixed $0< 
a<\frac{1}{2\gamma_+},$ as in \cite{EY2} we  define 

\begin{eqnarray}\label{b}
\hat b_{c_0,a,\beta}^\pm [g] &=&\int_0^L \int_{-a}^a \bigg(\bigg[ {(1 
\pm u\gamma_+)}^{-2} \pm \frac{1}{2} N_{c_0}(a)\bigg]|\partial_s g|^2  + 
|\partial_u g|^2 \nonumber \\
&+& \bigg[ - \frac{1}{4} \gamma(s)^2\pm \frac{1}{2}N_{c_0}(a) \pm 
M_{c_0}(a)   \bigg]|g|^2\bigg) duds \nonumber \\
&-& \beta \int_0^L|g(s,o)|^2ds - \gamma_+ b_\pm\int_0^L 
\bigg(|g(s,a)|^2 + |g(s,-a)|^2 \bigg)ds 
\end{eqnarray}

for $g\in {\cal{Q}} _a^\pm ,$ respectively. Since $|Im(\overline 
g\partial _sg)|\leq \frac{1}{2} \bigg(|g|^2 + |\partial_s g|^2\bigg),$ 
we obtain

\begin{eqnarray}\label{ac1}
\tilde b_{c_0,a,\beta}^+[g] \leq \hat b_{c_0,a,\beta}^+ [g] \qquad 
\mbox{for}\qquad \:  g\in {\cal{Q}} _a^+ 
\end{eqnarray}

\begin{eqnarray}\label{ac2}
\hat b_{c_0,a,\beta}^-[g] \leq \tilde b_{c_0,a,\beta}^- [g] \qquad 
\mbox{for}\qquad \:  g\in {\cal{Q}} _a^- .
\end{eqnarray}

Let $\hat H^\pm_{c_0,a,\beta}$ be the self-adjoint operators associated 
with the form 
$\hat b_{c_0,a,\beta}^\pm ,$ respectively.

Furthermore, let $T^+_{a,\beta}$ be the self-adjoint operator 
associated with the form

$t^+_{a,\beta}[f] = \int_{-a}^a |f'(u)|^2du - \beta|f(0)|^2; \qquad f\in H_0^1(]-a,a[),$ 

and similarly, let $T^-_{a,\beta}$ be the self-adjoint operator 
associated with the form

$t^-_{a,\beta}[f] = \int_{-a}^a |f'(u)|^2du - \beta|f(0)|^2- 
\gamma_+\big(|f(a)|^2+ |f(-a)|^2\big); \qquad f\in H^1(]-a,a[).$

As in \cite{EY2}, let us denote by $\mu_j^\pm(c_0,a)$ the $j^{th}$ 
eigenvalue of the following operator, define on $L^2(]0,L[)$, by 

\begin{eqnarray}
U^\pm_{a,\beta} = -\bigg[ {(1 \mp u\gamma_+)}^{-2} \pm \frac{1}{2} 
N_{c_0}(a)\bigg]\frac{d^2}{ds^2} - \frac{1}{4} \gamma(s)^2\pm 
\frac{1}{2}N_{c_0}(a) \pm M_{c_0}(a) 
\end{eqnarray}

in $L^{2}((0,L))$ with the domain $P_{c_0}$ specified in the
 previous section. Then we have
 \begin{equation} \label{decomp}
\hat{H}^{\pm}_{c_0,a,\beta}= U_{c_0,a}^{\pm}\otimes 1+1\otimes
T^{\pm}_{a,\beta}.
 \end{equation}
Let $\mu_{j}^{\pm}(c_0,a)$ be the $j$-th eigenvalue of
 $U^{\pm}_{c_0,a}$ counted with multiplicity. We shall prove the
 following estimate as in \cite{EY2}.
\begin{prop}\label{EST1}
Let $j\in \N.$ Then there exists $C(j)>0$ such that \\
$|\mu_j^+(c_0,a) -  \mu_j(c_0) | + | \mu_j^-(c_0,a) -  \mu_j(c_0) |\leq 
C(j)a$\\
holds for $c_0\in I$ and $0<a<\frac{1}{2\gamma_+},$ where $C(j)$ is 
independent of $c_0$ and $a.$
\end{prop}

{\bf{Proof:}}

{\sl Proof:} Since
 \begin{eqnarray*} 
\lefteqn{ U^{+}_{c_0,a} -\left\lbrack
(1-a\gamma_{+})^{-2}+\frac{1}{2}N_{c_0}(a) \right\rbrack S_{c_0}} \\
&& = \frac{1}{4}\left\lbrack
\frac{a\gamma_{+}(2-a\gamma_{+})}{(1-a\gamma_{+})^{2}}
+\frac{1}{2}N_{c_0}(a) \right\rbrack
\gamma(s)^{2}+\frac{1}{2}N_{c_0}(a)+M_{c_0}(a)\,,
 \end{eqnarray*}
 $N_{c_0}(a)+M_{c_0}(a)\leq Ta$ for $0<a<
\frac{1}{2\gamma_{+}}$ and $c_0\in I$, we infer that there is a
 constant $C_{1}>0$ such that
 $$ 
\left\Vert U^{+}_{c_0,a} - \left\lbrack
(1-a\gamma_{+})^{-2}+\frac{1}{2}N_{c_0}(a) \right\rbrack S_{c_0}
\right\Vert \leq C_{1}a
 $$ 
for $0<a<\frac{1}{2\gamma_{+}}$ and $c_0\in I$. This together with
 the min-max principle implies that
 $$ 
\left|\mu^{+}_{j}(c_0,a)- \left\lbrack
(1-a\gamma_{+})^{-2}+\frac{1}{2}N_{c_0}(a) \right\rbrack \mu_{j}(c_0)
\right | \leq C_{1}a
 $$ 
for $0<a<\frac{1}{2\gamma_{+}}$ and $c_0\in I$. Since
 $\mu_{j}(\cdot)$ is continuous, we claim that there exists a
 constant $C_{2}>0$ such that
 $$ 
\left|\mu^{+}_{j}(c_0,a)-\mu_{j}(c_0)\right| \leq C_{2}a
 $$ 
for $0<a<\frac{1}{2\gamma_{+}}$ and $c_0\in I$. In a similar way, we
 infer the existence of a constant $C_{3}>0$ such that
 $$ 
\left|\mu^{-}_{j}(c_0,a)-\mu_{j}(c_0)\right| \leq C_{3}a
 $$ 
for $0<a<\frac{1}{2\gamma_{+}}$ and $c_0\in I$.\\


Let us recall the following result from \cite{EY1}$\Box$.

\begin{prop}\label{EST2}
(a) Suppose that $\beta a>\frac{8}{3}.$ Then $T^+_{a,\beta}$ has only 
one negative eigenvalue, which we denote by $\zeta_{a,\beta}.$ It 
satisffies the ineqality\\
$\frac{-1}{4} \beta^2 < \zeta_{a,\beta} < \frac{-1}{4} \beta^2 + 
2\beta^2 \exp (\frac{-1}{2} \beta).$  \\
(b) Let $\beta >8$ and $\beta >\frac{8}{3}\gamma_+.$ Then 
$T_{a,\beta}^-$ has a unique negative eigenvalue $\zeta_{a,\beta}^-,$ and moreover, 
we have\\
$\frac{-1}{4} \beta^2 - \frac{2205}{16} \beta^2\exp (\frac{-1}{2} 
\beta)    < \zeta_{a,\beta} ^- < \frac{-1}{4} \beta^2 .$ 

\end{prop}

{\bf{Proof of theorem \ref{t1}}}

 We take
 $a(\beta) =6\beta^{-1}\ln\beta$. Let $\xi^{\pm}_{\beta,j}$ be
 the $j$-th eigenvalue of $T^{\pm}_{a(\beta),\beta}$, by
 Proposition~\ref{EST2} we have
 $$ 
\xi^{\pm}_{\beta,1}=\zeta^{\pm}_{a(\beta)\,,\beta},\quad
\xi_{\beta,2}^{\pm}\geq 0\,.
 $$ 
From the decompositions (\ref{decomp}) we infer that $\{
\xi^{\pm}_{\beta,j} +\mu_{k}^{\pm}
(B,a(\beta))\}_{j,k\in{\N}}$ properly ordered, is the
 sequence of the eigenvalues of $\hat{H}^{\pm}_{ c_0,a(\beta),\beta}$
 counted with multiplicity. Propositions~\ref{EST1} gives
 \begin{equation} \label{lowest}
\xi^{\pm}_{\beta,j}+\mu_{k}( c_0,a(\beta))\geq \mu_{1}^{\pm}(
c_0,a(\beta))=\mu_{1}(c_0)+\mathcal{O}(\beta^{-1}\ln\beta)
 \end{equation}
for $c_0\in I$, $j\geq 2$, and $k\geq 1$, where the error term is
 uniform with respect to $c_0\in I$. For a fixed $j\in \N$, we
 take
 $$ 
\tau^{\pm}_{c_0,\beta,j}=\zeta^{\pm}_{a(\beta),\beta}
+\mu_{j}^{\pm}(c_0,a(\beta)).
 $$ 
Combining Propositions~\ref{EST1} and \ref{EST2} we get
 \begin{equation} \label{tauasympt}
\tau^{\pm}_{c_0,\beta,j}=-\frac{1}{4}\beta^{2}
+\mu_{j}(c_0)+\mathcal{O}(\beta^{-1}\ln\beta)\quad\mathrm{as}
\quad\beta \to\infty\,,
 \end{equation}
where the error term is uniform with respect to $c_0\in I$. Let us
 fix $n\in\N$. Combining (\ref{lowest}) with
 (\ref{tauasympt}) we infer that there exists $\beta(n,I)>0$
 such that the inequalities
 $$ 
\tau^{+}_{c_0,\beta,n}<0,\quad
\tau^{+}_{c_0,\beta,n}<\xi^{+}_{\beta,j}+\mu_{k}^{+}(
c_0,a(\beta)),\quad
\tau^{-}_{c_0,\beta,n}<\xi^{-}_{\beta,j}+\mu_{k}^{-}( c_0,a(\beta))
 $$ 
hold for $c_0\in I$, $\beta\geq\beta(n,I)$, $j\geq 2$, and $k\geq
1$. Hence the $j$-th eigenvalue of $\hat{H}^{\pm}_{ c_0,a(\beta),
\beta}$ counted with multiplicity is $\tau^{\pm}_{c_0,\beta,j}$
 for $c_0\in I$, $j\leq n$, and $\beta\geq \beta(n,I)$. Let $c_0\in
I$ and $\beta\geq \beta(n,I)$. We denote by  $\kappa^{\pm}_{j}
(c_0,\beta)$ the $j$-th eigenvalue of $L^{\pm}_{c_0,a,\beta}$.
 Combining our basic estimate and the resultt of \cite{EY2} with
 Lemmas~\ref{1} and \ref{2}, relations
 \ref{ac1} and \ref{ac2}, and the min-max principle, we
 arrive at the inequalities
 \begin{equation} \label{finalest}
\tau^{-}_{c_0,\beta,j}\leq\kappa^{-}_{j}(c_0,\beta)\quad
\mathrm{and}\quad \kappa^{+}_{j}(c_0,\beta)\leq\tau^{+}_{
c_0,\beta,j}\quad\mathrm{for}\quad 1\leq j\leq n\,,
 \end{equation}
so we have $\kappa^{+}_{n}(c_0,\beta)<0<\inf
\sigma_{\mathrm{ess}}(H_{c_0,\beta})$. Hence the min-max principle
 and and the result of \cite{rs} imply that $H_{c_0,\beta}$ has at least $n$
 eigenvalues in $(-\infty,\kappa^{+}_{n}(c_0,\beta)]$. Given $1\leq
j\leq n$, we denote by $\lambda_{j}(c_0,\beta)$ the $j$-th
 eigenvalue of $H_{c_0,\beta}$. It satisfies
$$\kappa_{j}^{-}(c_0,\beta)\leq\lambda_{j} (c_0,\beta)
\leq\kappa_{j}^{+} (c_0,\beta)\quad\mathrm{for}\quad 1\leq j\leq
n\,;$$
this together with (\ref{tauasympt}) and (\ref{finalest})
 implies that
 $$ 
\lambda_{j}(c_0,\beta)=-\frac{1}{4}\beta^{2}
+\mu_{j}(c_0)+\mathcal{O}(\beta^{-1}\ln\beta)
\quad\mathrm{as}\quad\beta \to\infty\quad \mathrm{for} \quad
1\leq j\leq n\,,
 $$ 
where the error term is uniform with respect to $c_0\in I$. This
 completes the proof.

{\bf{Proof of corollary \ref{c1}}} 
The theorem\ref{t1} with \cite{rs} (theorem XIII.89)  yields the claim. 

\section{Remarks}

The essential of this paper is the determination of the unitary operator (\ref{up}) which permit us to have all the conditions of \cite{EY2}, to have our results.
 
{\bf Acknowledgments}

 The authors thank the Abdus Salam ICTP, the Belgian Cooperation CUD-CIUF-UAC/IMSP and the Conseil Régional Provence - Alpes - Côte d'Azur (France) for their financial support. They are thankful to Professor Jean-Michel Combes for fruitful collaboration.

\end{document}